# Violent suicide attempt history in advanced age individuals with bipolar disorder: the role of sex, abdominal obesity, and verbal memory: results from the FACE-BD cohort (FondaMental Advanced center of Expertise for Bipolar Disorders)

Aiste Lengvenyte[1,2,3], Bruno Aouizerate[2,4,5], Valerie Aubin[6], Joséphine Loftus[2,6], Emeline Marlinge[2,7], Raoul Belzeaux[2,8], Caroline Dubertret[2,9], Sebastien Gard[2,4], Emmanuel Haffen[2,10], Raymund Schwan[2,11], Pierre-Michel Llorca[2,12], Christine Passerieux[2,13], Paul Roux[2,13], Mircea Polosan[2,14], Bruno Etain[2,7], Marion Leboyer[2,15], FondaMental Advanced Centres of Expertise in Bipolar Disorders (FACE-BD) Collaborators; Philippe Courtet[1,2], Emilie Olié[1,2]

[1] Department of Emergency Psychiatry and Post-Acute Care, CHU Montpellier, France; IGF, University of Montpellier, CNRS, INSERM, Montpellier, France

[2] Fondation FondaMental, Créteil, France

[3] Faculty of Medicine, Institute of Clinical Medicine, Psychiatric Clinic, Vilnius University, Vilnius, Lithuania

[4] Charles-Perrens Hospital, Department of Clinical and Academic Psychiatry, Bordeaux

[5] France/NutriNeuro, UMR INRAE 1286, University of Bordeaux, Bordeaux, France

[6] Psychiatric Center, Hospital Princess Grace, Monaco, France

[7] AP-HP, GHU Paris Nord, DMU Neurosciences, Hôpital Fernand Widal, and INSERM UMRS 1144 and Université de Paris, Paris, France

[8] Pôle de Psychiatrie, Assistance Publique Hôpitaux de Marseille, Marseille, France; INT-UMR7289, CNRS Aix-Marseille Université, Marseille, France

[9] AP-HP, Groupe Hospitalo-Universitaire AP-HP Nord, DMU ESPRIT, service de Psychiatrie et Addictologie. Hopital Louis Mourier, Colombes, Inserm U1266, Faculté de médecine, Université de Paris, France

[10] Service de Psychiatrie de l'Adulte, CIC-1431 INSERM, CHU de Besançon, Laboratoire de Neurosciences, Université de Franche-Comté, UBFC

[11] Université de Lorraine; Centre Psychothérapique de Nancy, Pôle Hospitalo-Universitaire de Psychiatrie d'Adultes du Grand Nancy, Nancy, France; INSERM U1114

[12] Department of Psychiatry, CHU Clermont-Ferrand, University of Clermont Auvergne, EA 7280, Clermont-Ferrand, France

[13] Centre Hospitalier de Versailles, Service Universitaire de psychiatrie d'adulte et d'addictologie, Le Chesnay, Université Paris-Saclay, Université Versailles Saint-Quentin-En-Yvelines, DisAP-DevPsy-CESP, INSERM UMR1018, 94807, Villejuif, France

[14] Service Universitaire de Psychiatrie, CHU de Grenoble et des Alpes, Grenoble, France

[15] Univ Paris Est Créteil, INSERM, IMRB, Translational Neuropsychiatry, AP-HP, DMU IMPACT, FHU ADAPT, Fondation FondaMental, F-94010, Créteil, France

Corresponding author: Aiste Lengvenyte, MD
Email: aiste.lengvenyte@chu-montpellier.fr
Address: CHU Montpellier, 371 Av. du Doyen Gaston Giraud, 34090 Montpellier, France

*List of FondaMental Advanced Centre of Expertise (FACE-BD) collaborators is provided in the Acknowledgements section.

Abstract

**Background.** Bipolar disorder (BD) is a chronic, lifelong condition, associated with increased risk of obesity, cognitive impairment, and suicidal behaviors. Abdominal obesity and a higher risk of violent suicide attempt (SA) seem to be shared correlates with older age, BD, and male sex until middle age when menopause-related female body changes occur. This study aimed at assessing the role of abdominal obesity and cognition in the violent SA burden of individuals with BD.

**Methods.** From the well-defined nationwide cohort FACE-BD (FondaMental Advanced center of Expertise for Bipolar Disorders), we extracted data on 619 euthymic BD patients that were 50 years or older at inclusion. Cross-sectional clinical, cognitive, and metabolic assessments were performed. SA history was based on self-report.

**Results.** Violent SA, in contrast to non-violent and no SA, was associated with higher waist circumference, abdominal obesity and poorer California Verbal Learning Test short-delay free recall (CVLT-SDFR) (ANOVA, $p < .001$, $p = .014$, and $p = .006$). Waist circumference and abdominal obesity were associated with violent SA history independently of sex, BD type and anxiety disorder (Exp(B) 1.02, CI 1.00 – 1.05, $p = .018$; Exp(B) 2.16, CI 1.00 – 4.64, $p = .009$, accordingly). In an exploratory model, waist circumference and CVLT-SDFR performance mediated the association between male sex and violent SA.

**Limitations.** Cross-sectional design and retrospective reporting.

**Conclusions.** Violent SA history was associated with abdominal obesity and poorer verbal memory in older age BD patients. These factors were interlinked and might mediate the association between male sex and violent SA.

**Keywords:** Suicide; Bipolar disorder; Obesity; Cognitive function; Memory; Cohort

Background

Bipolar disorder (BD) is a debilitating, lifelong illness with lifetime prevalence of up to 2.5% (Merikangas et al., 2011). In addition to a two-fold increased mortality related to chronic cardiovascular and metabolic diseases (Crump et al., 2013), individuals with BD die by suicide 20 to 30 times more frequently than their peers in general population (Crump et al., 2013; L. Plans et al., 2019). High rates of violent method use for suicide attempts (SA) confer to this excess mortality in BD (Laura Plans et al., 2019). Individuals with BD are also the most likely to engage in repeated SA among all psychiatric disorders (Mora et al., 2017). Generally, SA risk is considered the most elevated at the young age, and most studies exploring the risk factors for SA were dominated by these attempts in early age. However, suicide risk increases among BD patients with a longer duration of illness (Altamura et al., 2010), underscoring the need to investigate risk factors in older age patients.

The most consistent and robust risk factors for suicide in BD is a personal history of SA (Miller and Black, 2020). In this regard, violent SA are of particular importance, as they are not only associated with an increased risk of handicap and death, but are also related to a three-fold increased risk to repeat SA, frequently also entailing higher lethality (Giner et al., 2014). Variables that are the most robustly associated with violent SA in both general population and individuals with BD are male sex and older age (Giner et al., 2014; Nivoli et al., 2011). Notably, another shared trait among BD, older age, male gender and people that died from a violent SA is increased levels of pro-inflammatory markers (Bakker et al., 2018; Fernandes et al., 2016; Gananca et al., 2016; Villeda et al., 2011).

Over half of individuals with BD become obese within 20 years after their first hospitalization – twice the rate of the general population (Strassnig et al., 2017). Studies in BD revealed positive relationships between obesity and SA (Gomes et al., 2010), and violent

SA (Rosso et al., 2020). Obesity has been associated with worse global cognitive ability and poorer verbal memory in BD (Depp et al., 2014; Mora et al., 2017). Increased body mass index has been associated with white structure abnormalities in tracts that are crucial to mood regulation and neurocognitive functioning, suggesting that obesity might contribute to the pathophysiology of BD through a detrimental action on structural connectivity in cortico-limbic networks (Mazza et al., 2017). Meanwhile, poorer subjective cognitive functioning has been associated with both - BD diagnosis, and SI in BD patients (Luo et al., 2020)

Several neurocognitive variables, such as worse performance in verbal learning and impulse control tasks, have been associated with violent SA history, suggesting a neurocognitive vulnerability (Lengvenyte et al., 2019). Studies repeatedly show that suicide attempters have poorer verbal memory performance, with the highest effect in violent suicide attempters (Perrain et al., 2021). Meanwhile, patients with a violent SA history have been reported to have higher cerebrospinal fluid insulin than those with nonviolent SA history (Westling et al., 2004). Indeed, multiple studies have reported and association between poorer verbal memory with abdominal obesity and insulin resistance in BD (Isaac et al., 2011; Lackner et al., 2016; Salvi et al., 2020), suggesting the role of cognitive function in the relationship between obesity and violent SA.

With regard to sex, males with BD have been reported to engage in more lethal SA compared to both males with major depressive disorder and females with BD (Zalsman et al., 2006). In a sample of recent suicide attempters, only male gender, diagnosis of BD, lower total cholesterol, and higher C-reactive protein levels in serum predicted high-lethality SA (Aguglia et al., 2019). It should be noted that males accumulate more visceral, while females have more subcutaneous fat in the first half of life, offering the latter estrogen-related protection from the negative consequences related to obesity. This suggest that male-related phenotype with higher accumulation of abdominal fat might be implicated in increased suicide risk in this sex. This female advantage, however, is lost at around 50 years of age, when they undergo the menopausal transition, characterized by an increased testosterone to estrogen ratio and related abdominal adipose tissue depots (Goossens et al., 2021).

Yet, little is known regarding the role of abdominal obesity, which is a major component of metabolic syndrome known to be twice more frequent in BD than in the general population (Godin et al., 2014). Waist circumference correlates with increased activation in the frontal lobes during a mental stress task (Moazzami et al., 2020) – suggesting an increase in required effort to coordinate a response to a stressful situation, putting the individual at risk to miscalculate when the stressor is excessive or prolonged. Mechanistically, visceral adipose tissue, in contrast to the subcutaneous adipose tissue, is more metabolically and inflammatory active. When expanded, visceral tissue produces pro-inflammatory cytokines that further sustain the low-grade inflammation (Schipper et al., 2012). The blood-brain barrier damage in BD, known to be associated with insulin resistance, might facilitate the transition of these proinflammatory molecules to the brain (Kamintsky et al., 2020). Decreased coupling strength between structural and functional connectome in BD (Zhang et al., 2019) may further contribute to the brain function impairment in response to these pro-inflammatory molecules

To the best of our knowledge, no study has specifically investigated the association between abdominal obesity, cognitive function, SA and its violence in individuals with BD. To address this knowledge gap, we assessed the association between SA types and measures of obesity in individuals with BD over 50 years of age, when the protective period in females is lost. In addition, we analyzed patient performance in an extensive cognitive battery, hypothesizing that a certain cognitive impairment might mediate the association between abdominal obesity and SA. Finally, we assessed the role of these factors in the well-established association between male sex and violent SA.

## Methods

### Patient evaluation

This FACE-BD (FondaMental Advanced Center of Expertise for Bipolar Disorders) is a multicentre cross-sectional study where patients are recruited by a French national network of nine BD expert centres that were created to follow and evaluate BD patients in a structured and comprehensive way (Henry et al., 2011) under the coordination of FondaMental foundation (www.fondation-fondamental.fr). Patients are assessed by a senior psychiatrist and clinical psychologist. The assessment protocol was approved by an ethical review board (Comité de Protection des Personnes Ile de France IX, 18 January 2010) in accordance with French law for non-interventional studies. Although the committee waived the requirement for written informed consent, the patients received a letter informing them of the study and for consent to participate in the research evaluation.

Participants sociodemographic, disease (type of BD, date of onset, history of rapid cycling), treatment (lifetime medication history) characteristics, as well as family history mood disorders and SB history collected obtained during the clinical interview. Experienced psychiatrists used the Structured Clinical Interview for DSM-IV Axis I Disorders (First, 1996), to confirm the diagnosis of BD and to assess the lifetime history of mood episodes, course and comorbid psychiatric disorders according to the Diagnostic and Statistical Manual, Fourth Edition (DSM-IV-TR) criteria, as this manual was the most widely accepted diagnostic tool in 2011, when the beginning of the recruitment of the patient cohort started. Outpatients with confirmed BD (type I, type II or not-otherwise-specified) aged 50 years or more at the time of assessment were eligible. Due to the cognitive evaluation, patients with a history of dyslexia, dysorthographia, dyscalculia, dyspraxia or language delay, and a history electroconvulsive therapy in the past year were excluded. Patients with any other type of comorbidity were eligible, and the most prevalent and systemically evaluated comorbidities (any type of anxiety disorder and substance use disorder) were included in the analysis.

Somatic comorbidity is systematically screened by self-reported "yes-or-no" questions, and the presence of diabetes and any neurological comorbidly (migraine, epilepsy, major head trauma or vascular event history, multiple sclerosis, and other neurological disorder previously established by a neurologist) were included in the current analysis. Since we evaluated the cognitive function that might be influenced by mood episodes, patients with significant mood symptoms (Montgomery-Asberg Depression Rating Scale (MADRS) (Montgomery and Asberg, 1979) score >19, Young Mania Rating Scale (YMRS) (Young et al., 1978) score $>12$) were excluded. Scores of these scales were used to measure the residual depressive and manic symptomatology. Sleep quality was evaluated using the Pittsburg Sleep Quality Index (PSQI) (Buysse et al., 1989). Childhood adversity was evaluated using the Childhood Trauma Questionnaire (CTQ) (Bernstein et al., 2003). Affect intensity was evaluated using the Affect Intensity Measure (AIM) (Rubin et al., 2012). Higher scores indicated higher severity in all cases.

Suicide attempt history was based on a self-report question "Have you ever attempted suicide?", which was followed by the interrogation regarding the time and type of the attempt. Violent suicide attempters were described as patients that reported at least one violent SA, which was conceptualized according to Asberg's criteria (hanging, the use of firearms, jumping from heights, several deep cuts, car crash, burning, gas poisoning, drowning, electrocution, and jumping under a train), as described earlier (Giner et al., 2014). Non-violent attempters were conceptualised as all other suicide attempters.

Metabolic health factors

Patients received a medical examination by a healthcare professional. Body weight was measured fasting and undressed, height barefoot, waist circumference at midway between the inferior margin of the ribs and the superior border of the iliac crest, at minimal inspiration. A blood pressure measurement was obtained by using a mercury sphygmomanometer in a seated position. A blood sample (for glucose (Glu), triglycerides (Tg) and high-density cholesterol (HDL-Chol)) was drawn in the morning, after overnight fast. The metabolic syndrome was defined according to the criteria of the International Diabetes Federation (IDF) (Alberti et al., 2006), which requires the presence of three or more of five criteria: (1) fasting glycemia ≥ 5.6 mmol/l or glucose-lowering medication, (2) HDL cholesterol < 1,03 mmol/l in males and < 1,29 mmol/l in females or history of specific treatment for this lipid abnormality, (3) triglycerides ≥1.7 mmol/l or on lipid-lowering medication, (4) systolic arterial blood pressure ≥130 mmHg or diastolic arterial blood pressure ≥85 mmHg or antihypertensive treatment, (5) umbilical perimeter ≥94 cm for men and ≥80 cm for women. Lipid-based and glycemia-based insulin resistance indicators were counted by dividing Tg/HDL-Chol and Glu/HDL-Chol, accordingly, and sex-specific reference measures were used to assess resistance (Watson et al., 2020; Young et al., 2019).

For further comparisons, in response to an elevated age and mean waist circumference of the sample, more liberal waist circumference thresholds by were utilized (≥88 cm for females and ≥102 cm for males), according to the American National Institute of Health guidelines (Ross et al., 2020). For further calculations, we then categorized patients according to their sex-based waist circumference in each BMI category, as suggested by the recent endocrinologists' consensus, qualifying waist circumference as excessive when ≥80 cm in females and ≥ 90 in males in normal weight (BMI < 25 kg/m$^2$), ≥90 cm and ≥100 cm in overweight (BMI 25 – 29.9 kg/m$^2$), ≥ 105 cm and ≥110 cm in moderately obese (BMI 30 – 34.5 kg/m$^2$), and ≥115 cm and ≥125 cm in severely obese (BMI ≥ 35 kg/m$^2$) individuals (Ross et al., 2020).

Cognitive battery

Patients were administered a standardized cognitive test battery, which complied with the recommendations by the International Society for Bipolar Disorders (Yatham et al., 2010). Twelve tests evaluating three cognitive domains were included in the present study: (1) the digit symbol coding and symbol search subtests from the Wechsler Adult Intelligence Scale (WAIS) version III (Wechsler, 1997), the Trail Making Test (TMT) part A (Llinàs-Reglà et al., 2017), and the word and the color conditions of the Stroop test (Van der Elst et al., 2006) to evaluate processing speed;(2)The color/word condition of the Stroop test, the TMT part B and verbal fluency (semantic and phonemic conditions) (Lezak, 2004) for executive functions; and the California Verbal Learning Test (CVLT) immediate recall, short and long delay free recall (Delis D, Kramer J, Kaplan E, 2000) for verbal memory. Raw scores were transformed to demographically corrected standardized (z) scores based on normative data for each test (Godefroy, 2012; Golden, 1978; Poitrenaud et al., 2007; Wechsler, 1997; Wechsler et al., 2008) and then transformed to T scores expressed in percentiles where 50 corresponds to 0 of the z scores. Higher scores reflected better performance. Cognitive data obtained using this battery have been published previously (Roux et al., 2019).

Statistical analysis

Initial univariate comparisons were performed with $\chi^2$ test for categorical variables and ANOVA or Kruskal-Wallis test continuous variables, in accordance to distribution normality. Post-hoc Tukey honestly significant difference (HSD) test was used for homogenic variables, and Games-Howell test was used for non-parametric. Before comparisons, continuous data was assessed for normality and homogeneity with the Shapiro-Wilk and Levene's tests. Winsorizing procedure was applied for outliers.

Due to the specify of the data collection process, there were multiple missing values. In all cases, it was established that data was missing at random Little's missing completely at random test p >.1. Pairwise complete case method was used for comparisons. Number and % of missing values for each variable are reported in Table 1 and Table 2.

We then compared violent attempters to non-violent attempters and both non-violent attempters and non-attempters on the three terciles-based ranges of waist circumference, and examined with linear logistic regression how these ranges predicted violent suicide attempt, controlled for sex. Equivalent procedure was done in females, with female waist circumference terciles. Estimated marginal means (EMM) with 95% confidence intervals [95% CI] are reported. Odds ratios (OR) and the receiver operating characteristic curves (ROC) were assessed to describe the predictive properties of the significant variables. We then performed multiple logistic regression to see the independence between abdominal obesity and violent SA. Model 1 controlled for sex and BD type. Model 2 controlled for variables in model 1 and anxiety disorder history. Model 3 was performed in females only and controlled for BD type. To assess performance in the prediction of violent suicide attempt, we calculated the Hosmer– Lemeshow and Nagelkerke's generalized model $R^2$ statistics. The multicollinearity was checked with the variance of inflation factor (VIF), and collinearity was refused if VIF <3. OR are expressed in Exp(B).

Possible mediation effects were analyzed using the SPSS PROCESS macro (Preacher and Hayes, 2008) with 1000 iterations Bootstrapping. It has been shown to provide reliable results in previous neurobiology research (Opel et al., 2019). PROCESS estimates direct and indirect effects between variables by applying an ordinary least squares path analytic framework. For dichotomous outcome variables, logistic regression models were applied. To test inference of indirect (mediated) effects, bootstrap confidence intervals were calculated. P values for indirect effects are not reported in PROCESS, and significance of indirect effects is assumed if the 95% CI does not include zero. To test our hypothesis, a mediation model included sex as an independent variable, waist circumference and CVLT short delay free recall as mediators, and type of suicide attempt (violent vs non-violent) as a dichotomous outcome variable. Subsequently, two additional models that used sex as a nuance variable to examine the role of CVLT short delay free recall in the association between waist circumference and violent SA. They corrected the effect distribution between multiple pathways. Unstandardized regression coefficients and standard errors (SE) are presented for each effect.

We deemed 2-sided p values < .05 statistically significant in multivariate comparisons and Benjamin-Hochberg False Discovery Rate (FDR)-corrected p values <.006 in initial univariate comparisons (N = 43). SPSS (IBM) version 25 was used for all analyses.

Results

3.1. Sociodemographic and clinical characteristics

The total sample included N=619 individuals with BD aged 50 years or more. 59.3% (N = 367) of patients were females, and had 57.69 (SD 6.42) years on average. Two thirds (N=344) of patients were married of living in a couple and patients had 14.01 (SD 3.25) years of education. 38.3% (N = 237) of patients had BD type 1. Average illness onset was at 30.84 (SD 12.63) years, with the mean duration of 26.90 (SD 12.77) years. 16% (N = 85) had rapid cycling type of disorder. Some type of neurological comorbidity (migraine, epilepsy, multiple sclerosis, meningitis, severe head trauma, and cerebrovascular event history) was present in 31.5% (N = 187) cases. Finally, 37.6% (N = 233) of patients had a lifetime history of SA, and in 23.4% (N = 54) cases it was violent. More detailed sociodemographic and clinical characteristics stratified by suicide attempt history and type (violence) are presented in Table 1. Notably, percentages in the text are based on number of patients excluding those with the missing value for the evaluated variable, and more detailed sociodemographic and clinical characteristics stratified by suicide attempt history and type (violence), as well as the numbers of missing values are presented in Table 1.

All three groups had comparable age, educational attainment, marital status, family history of mental illnesses and suicide, as well as treatment history. We observed an FDR-corrected significant association between non-violent SA history and female sex, as compared to both violent attempters and non-attempters (78.0% vs 53.7% vs 51.5 %, accordingly, $p < .001$). Substance use disorder (SUD) history and rapid cycling history frequency were comparable in violent and non-violent suicide attempters, and significantly exceeded the observed rates in non-attempters, though only the first comparison survived the FDS correction (31.4% vs 32.5% vs 16.4%, $p < .001$, and 24.4% vs 20.1% vs 13.1%, $p = .047$, accordingly). Compared to non-attempters, non-violent suicide attempters also had earlier illness onset and longer duration, as well as higher affect intensity (AIM scores 3.68 (SD .60) vs 3.51 (SD .62), Tukey HSD $p = .010$), and sleep impairment (PSQI scores 7.93 (SD 3.53) vs 6.79 (SD 3.54), Games-Howell $p = .002$). There was a tendency for lower PSQI score in violent attempters compared to non-violent attempters as well, but it did not reach the level of statistical significance (Games-Howell $p > .05$). Tendencies for associations between type I BD and a violent SA, and between anxiety disorder history and non-violent SA were also observed, but these associations did not survive FDR correction (p >.006).

Table 1. Sociodemographic and clinical characteristics stratified by suicide attempt history and violence

3.2. Metabolic health, cognitive factors and suicide attempts

Among metabolic health factors, only one association survived the FDR correction – association between waist circumference continuous measure and SA history, with violent attempters having the highest waist circumference, and non-violent attempters the lowest, and both violent attempters and non-attempters exceeding non-violent attempters [101.43 (SD 12.75) cm in violent attempters vs 97.17 (SD 14.26) cm in non-attempters vs 92.89 (SD 12.82) cm in non-violent attempters, p < .001]. Further analyses using sex-specific cut-off points revealed a specific association between violent SA history and both abdominal obesity and waist fat excess in each BMI group, as compared to both non-violent attempters and non-attempters, but these associations did not survive FDR correction. No associations between suicide attempter status and type were found with glycemia-based or lipid-based presumed insulin resistance and metabolic syndrome.

Patients in the study sample had the tendency underperform in most cognitive tasks except Digit Symbol Coding and Digit Symbol Search compared to expected scores based on their age and sex. Regarding cognitive functions, violent attempters performed worse on all verbal memory tasks, as well as in in Stroop word condition and Phonemic verbal fluency task, though only the association between lower performance in CVLT short delay free recall task performance and violent suicide attempt history survived FDR correction (42.49 (SD

10.94) in violent attempters vs 47.29 (SD 13.27) in non-attempters vs 49.09 (SD 11.38) in non-violent attempters, p = .006).

Table 2. Metabolic and cognitive factors stratified by suicide attempt history and violence

3.3. Specific association between visceral obesity and violent SA

To further test association between waist circumference and violent SA history, we performed logistic regression models with waist circumference terciles (< 90 cm, 90-101.9 cm, and ≥102 cm) as predictor and violent SA as a binary outcome. Since sex is strongly associated with both waist circumference and violent SA history, biological sex was included as a nuisance covariate in the models. When adjusted for sex, the probability to have attempter violent SA increased from the lowest to the middle and the highest terciles (EMM .019, [95% CI -.004 – .044] vs .122, [95% CI .077 - .172] vs .107, [95% CI .065 - .154], p = .001, effect of sex =.40).

Similarly, such effect was observed in suicide attempters (EMM .079, [95% CI .023 – .147] vs .299, [95% CI .198 - .412] vs .321, [95% CI .202 - .442], p = .001, effect of sex =.27). To further test if the association is independent of sex, the analyses were performed only in females, using female waist circumference terciles (< 85 cm, 85-96.9 cm, and ≥97 cm). The associations remained statistically significant with regard to all females (EMM .019, [95% CI -.032 - .049] vs .094, [95% CI .042 - .153] vs .111 [95% CI .054 - .167] in the lowest, middle and highest percentiles, p = .025) and suicide attempters only (EMM .043, [95% CI -.065 - .115] vs .216 [95% CI .109 - .327] vs .255 [95% CI .141 - .391] in the lowest, middle and highest percentiles, p = .013). The relationship between either waist circumference continuously or its terciles was not significant in males. The probability of violent SA according to waist circumference terciles is presented in Figure 1.

Figure 1. Probability of suicide attempt and violent suicide attempt in all patients, suicide attempters, and female suicide attempters

Among suicide attempters, ROC area under the curve (ROC-AUC) for abdominal obesity was: .583, SE .040, 95% CI .505 - .661, p = .043. Likewise, significant ROC-AUC were detected for W waist circumference terciles and violent SA in all suicide attempters (ROC-AUC .368, SE .039, 95% CI .292 - .444, p = .002) and in female attempters alike (ROC-AUC .396, SE .040, 95% CI .317 - .475, p = .016).

To further test the association between the violent SA and obesity-related factors, we applied several multiple logistic regression models with obesity-related variables and other sociodemographic and clinical variables among those separating violent from non-violent SA attempters, namely, sex, BD type (type 1 vs other), and anxiety disorder (Table 1 and Table 2). Four possible measures related to fat accumulation on waist were entered in models: (1) waist circumference as a continuous variable; (2) waist circumference terciles; (3) sex-specific abdominal obesity; (4) sex-specific waist excess. A binary multiple logistic regression model controlled for sex and BD type (Model 1) revealed a statistically significant association between the first three variables and violent SA as compared to both non-violent attempters and non-violent attempters and non-attempters as a group. The Exp(B) for violent SA history in patients with abdominal obesity was 3.256 [95% CI 1.455 -7.287], p = .005 in suicide attempters, and 2.579, [1.273 – 5.225] in all patients p = .006.

A second model, which also included anxiety disorder history as a covariate, showed comparable results. The Exp(B) for violent SA history in patients with abdominal obesity

was 3.029, [95% CI 1.288 – 7.127], p = .005 in suicide attempters, and 2.157, [95% CI 1.003 – 4.641], p = .009 in all patients. Finally, a separate model in females which controlled for BD type also yielded significant results for waist circumference as continuous variable and as sex-based terciles, and abdominal obesity. The Exp(B) in females with abdominal obesity for violent SA history was 5.184, [95% CI 1.466 – 18.329], p = .005 in suicide attempters, and 5.374, [95% CI 1.574 – 18.351], p = .005 in all females. (see Table 3).

Table 3. Multiple logistic regression models for the association between visceral adiposity measures and violent SA history

3.4. Exploration of the possible direct and indirect relationships between sex, waist circumference, verbal memory and suicide attempt violence

The exploratory mediation model (N = 173) confirmed the significant association between sex and waist circumference (coefficient 9.496, SE 2.159, 95% CI 5.234 to 13.757, p < .001) and CVLT short delay free recall (coefficient −4.943, SE 1.859, 95% CI −8.612 to −1.274, p = .009). Likewise, significant associations in the mediation model were observed between waist circumference (coefficient .038, SE .015, 95% CI .008 to .067, Z = 2.486, p = .013; figure 2, path d) and CVLT short delay free recall (coefficient −.040, SE .0172, 95% CI −.0737 to −.006, Z = −2.321, p = .020; path c) with violent SA. Additional model was used examine the role of CVLT short term free recall in the association between waist circumference and violent SA. When controlled for sex, waist circumference was associated with CVLT short delay free recall (coefficient −.144, SE .065, 95% CI −.273 to −.016, p = .028, path f). The direct association between sex and CVLT short delay free recall became insignificant (coefficient -3.575, SE 1.939, (95% CI -7.403 to .253, p = .067, path a). In an identical inverse model, sex direct association with waist circumference remained significant (coefficient 8.535, SE 2.179, (95% CI 4.235 to 12.835, p < .001, path b). The general mediation model yielded a significant positive mediated effect of male sex on violent SA history through waist circumference and CVLT short-delay free recall (indirect effect: coefficient .556, SE .227, 95% CI .211 to 1.113; path e'), while direct effect was no longer significant (coefficient .532, SE .400, 95% CI -.252 to 1.316, Z = 1.331, p = .183; path e), indicating a significant mediation effect of sex on violent SA history through waist circumference and short-term verbal memory.

Figure 2. The applied mediation model between sex and violent suicide attempt via waist circumference and short-term verbal memory.

Discussion

In a cohort of 619 individuals with BD that are over 50 years old, we report a relationship between abdominal obesity and violent SA history, independently of sex, BD type, and anxiety. We also observed that increased weight circumference in old age could mediate the association between male sex and violent SA history, and that poor verbal memory is possibly involved in this relationship.

We report significantly (FDR-corrected) higher waist circumference in individuals with violent SA history compared to non-violent suicide attempters. Notably, both abdominal obesity and violent SA rates are particularly high in BD (Fagiolini et al., 2005; Laura Plans et al., 2019). In non-corrected comparisons, violent suicide attempters also exceeded all other in prevalence of sex-based cut-off values-based abdominal obesity and waist excess. Several logistic regression models with sex, BD type and anxiety disorder history as covariates further supported these results, showing a relationship between waist circumference, sex-based abdominal obesity and violent SA history. In agreement with another study in BD (D'Ambrosio et al., 2012), we found no association between SA and metabolic syndrome or several proxy measures for insulin resistance (Watson et al., 2020; Young et al., 2019), suggesting a specific role of abdominal fat depots, which is independent of metabolic syndrome and insulin resistance. However, while our results confirm the lack of association between the concentration of high-density cholesterol and violent SA history observed in previous studies, we did not measure the low-density cholesterol and very low-density cholesterol levels that have been associated with violent suicide attempt in patients with mood disorders (Capuzzi et al., 2020)

Allostatic load, a dysregulated stress-response system and abdominal obesity-related chronic low-grade inflammation might underlie the relationship between abdominal obesity and violent SA in BD. Generally, BD patients have elevated levels of pro-inflammatory cytokines, kynurenine metabolites, and awakening cortisol, even in euthymia (Carvalho et al., 2020). Elevated kynurenine to tryptophan ratio and cortisol levels are associated with increased reward-driven comfort food intake, which is tightly linked to with abdominal fat depots (Dalkner et al., 2018; Klatzkin et al., 2019). Meanwhile, abdominal obesity is linked to increased hypothalamus-pituitary-adrenal (HPA) axis responsivity, and increased expression of glucocorticoid receptors in adipose tissue (Incollingo Rodriguez et al., 2015). Glucocorticoids increase adipogenesis preferentially in visceral fat depot (Lee et al., 2014), possibly further sustaining the vicious cycle of obesity and impaired stress response. Abdominal adipose tissue expansion promotes low-grade inflammation via local production of pro-inflammatory cytokines (Schipper et al., 2012). Higher baseline cerebrospinal fluid pro-inflammatory cytokine IL-6 and plasma cortisol levels, as well as HPA axis hyper-reactivity in to a stress challenge have been associated with a history of violent SA (Adrián

Alacreu-Crespo et al., 2020; Isung et al., 2014). In support, concomitant obsessive compulsive disorder, which has robust associations with low-grade neuroinflammation (Attwells et al., 2017), has also been associated with the use of violent methods for SA in BD patients (Di Salvo et al., 2020).

Not surprisingly, male sex was the factor the most strongly associated with the choice of violent SA means – a well-accepted association in any patient group, including BD (Giner et al., 2014; Nivoli et al., 2011). As commonly accepted, males had slightly higher BMI and much larger waist circumference measures – the latter mediated the association between male sex and violent SA history. It should be noted that sex-specific waist circumference thresholds for obesity in white adults were originally developed based on average measurements in individuals that had a BMI of $30.0 \, kg/m^2$, and not health or functioning related outcomes (Janssen et al., 2002). waist circumference, and not BMI, is associated to cardiovascular disease (Fan et al., 2016), and is much more related to overall mortality due to any cause (Ross et al., 2020). Indeed, in spite of this accepted *normalization* of visceral fatness in males, they suffer from significantly higher cardiovascular disease burden and related lower life expectancy (Van Oyen et al., 2013). The present study adds violent SA as another factor that might contribute to higher mortality in males, in response to their higher weight circumference. which might be related to increased androgen load known to be associated to suicide completion (Lenz et al., 2019). Several studies also showed increased expression of androgen receptors in people with BD (Owens et al., 2019) and high levels of androgen receptors in visceral adipose tissue (Ibrahim, 2010), suggesting an interaction effect.

Finally, impaired performance in a short-term verbal memory was the cognitive variable the most strongly associated with violent SA, and the only one that survived FDR correction. Non-violent SA history was associated with the best performance on this task, echoing to a previous study that has related better verbal memory and suicidal behaviors (that are dominated by non-violent SA) (McHugh et al., 2021). Memory formation is important for successful decision-making process (A. Alacreu-Crespo et al., 2020), and poorer verbal memory performance in suicide attempters has been demonstrated to be particularly marked in violent suicide attempters (Perrain et al., 2021).

Poorer short-term verbal memory was also associated with higher waist circumference, and mediated the association between it and violent SA. Our model agrees to and extends on the previous observations of poorer verbal memory association with obesity and male sex in BD (Carrus et al., 2010; Mora et al., 2017). In contrast, higher verbal memory performances in females was observed in our sample, as partly supported by a previous study showing the memory advantage only in females with low pro-inflammatory cytokine IL-1β levels, further implicating the inflammatory pathway (Caldwell et al., 2020). Persistently elevated cortisol, which is frequently observed in individuals with abdominal obesity, has also been associated with poorer verbal memory (Segerstrom et al., 2016).

Lower hippocampal volume and exacerbation of the negative effect of aging on memory network covariance have been implicated in the relationship between abdominal adiposity and poorer memory (Isaac et al., 2011; Zsido et al., 2019). Our findings are also in line with recent preclinical findings. Obese subjects have increased levels of the lipid messenger prostaglandin $E_2$ ($PGE_2$), a major modulator of inflammation (García-Alonso et al., 2016). $PGE_2$ signaling has been recently associated with impaired bioenergetics of the ageing myeloid cells, maladaptive pro-inflammatory responses, lower synaptic plasticity, and impaired spatial memory in mice (Minhas et al., 2021), it is worth mentioning that the anti-inflammatory drug infliximab increased performance in a verbal memory task and improved anhedonia in a small trial with BD patients (Lee et al., 2020; Mansur et al., 2020). Lithium, an agent with proven long-term effectiveness for preventing suicide in BD, has also been demonstrated to reduce pro-inflammatory cytokine levels and improve verbal memory in BD (Burdick et al., 2020; Del Matto et al., 2020; Nassar and Azab, 2014).

Regarding treatment, it should also be noted that the observed relationship could reflect failed treatment response or adherence in individuals with violent SA history. Abdominal obesity, but not metabolic syndrome, has been associated with less improvement after 6 months of lithium- or quetiapine-based treatment in BD patients (Mcelroy et al., 2016). On the other hand, patients with more severe illness course are more likely to receive complex pharmacotherapy, generally including second generation antipsychotics (Kim et al., 2021), that might contribute to increased visceral adiposity (Nicol et al., 2018).

Meanwhile, it should be noted that other major factor not explored in the present study is the role of abdominal obesity in functioning and social status in patients. Increased weight circumference has been associated with worse interpersonal functioning (Yusufov et al., 2021). Weight stigma has been associated with anxiety, perceived stress, antisocial behavior, substance use, coping strategies, social support, and lower adherence to treatment, (Papadopoulos and Brennan, 2015). Independent of abdominal fat, weight stigma has been significantly related to several measures of cortisol as well as higher levels of oxidative stress, and perceived stress mediated the relationship between weight stigma consciousness and the cortisol awakening response in a previous study (Tomiyama et al., 2014). This might cause a vicious cycle, as perceived stress has been associated with increased weight circumference and levels of inflammatory markers (Gowey et al., 2019), resulting in self-sustaining obesity-stress cycle.

### Strenghts and limitations

The major strengths of the present study include the well-defined sample of a large number of older age individuals with BD, and use of multiple validated tests. Nevertheless, the results should be interpreted in relation the study limitations. First, the exploratory cross-sectional design hinders the conclusions regarding causality, most notably making the mediation model only exploratory. Second, the use of the DSM-IV-TR diagnostic manual, which was chosen due to its prevalence at the time of the cohort establishment, might miss some patients currently diagnosed with BD when using the current DSM-V diagnostic manual. A bidirectional or a multidirectional relationship is possible, and we cannot be completely confident that individuals were obese before attempting suicide - this relationship is yet to be proven in future longitudinal studies. Furthermore, our sample might not be representative of all older age patients with BD, as only a certain group of patients might be more likely to be referred to the expert center (possibly excluding those patients that are chronically hospitalized or failing to reach relative remission, as well as those with a favorable course of illness, which do not see the interest of an expert evaluation). In addition, we analyzed lifetime SA, but the sample does not include completed suicides – suggesting that we could have studied the "survivor cohort", where only patients that did not die during their SA are examined. Other factor to be considered. Nevertheless, violent SA history is the major risk factor for suicide reattempts, and understanding factors conferring to this increased risk is clinically meaningful. Retrospective reporting might have caused recall bias.

Another limitation is that the limited number of violent attempters did not allow us to control for all possible covariates. While we did not find an association with the use of any group of medications, some agents are notoriously known for their metabolic effects and future studies should further look into it. Unfortunately, due to the limited sample size we could not evaluate the role of each medication and their dose separately, which represents additional limitation of the present study. Medication use might also be confounded by disease severity, as more severe patients are more likely to receive complex pharmacotherapy, generally including second generation antipsychotics (Kim et al., 2021).

. Furthermore, we did not include comorbid medical and some psychiatric conditions (such as eating disorder history) that might contribute to suicide mortality, changed metabolic profile, and cognitive function, and cannot rule out that a complex interaction with multiple clinical and subclinical condition might confound the observed relationships. The role of such key suicide- and obesity-related factors as lifestyle factors, early life adversity, impulsivity and sleep quality are to be further examined in larger studies. We also used waist circumference as a proxy for visceral fatness, which is clinically meaningful but is not as specific as the visceral adipose tissue mass (Ross et al., 2020). While we included individuals over 50 years of age to cover the period when hormonal differences are marked, we did not collect data regarding female menopause and hormonotherapy status.

## Conclusion

In conclusion, our results indicate the possible role of abdominal obesity in increased risk of violent SA in older age BD patients - and that verbal memory impairment might be implicated in this relationship. The cross-sectional nature of our study does not allow us to establish causality, for which prospective hypothesis-based studies are needed. Given the observed relatively specific association between abdominal obesity and violent SA, interventional studies aiming at decreasing abdominal obesity in individuals with BD that measure SA or SI as an outcome would be of clinical and scientific value.

Table 1. Sociodemographic and clinical characteristics stratified by suicide attempt history and violence

| Characteristic | VSA (N=54) Mean (SD) or N (%) | NVSA (N=177) Mean (SD) or N (%) | No-SA (N=388) Mean (SD) or N (%) | P-value | Post-hoc | Missing No. (%) |
|---|---|---|---|---|---|---|
| **Sociodemographic characteristics** | | | | | | |
| Sex, female | 29 (53.7%) | 138 (78.0%) | 200 (51.5%) | <.001** | NVSA > VSA, No-SA | - |
| Age, years‡ | 57.56 (5.96) | 57.49 (5.98) | 57.80 (6.69) | .994 | .. | - |
| Education, years‡ | 14.29 (3.12) | 14.04 (2.82) | 13.95 (3.45) | .785 | .. | 95 (15.3%) |
| Married/ in couple | 27 (60.0%) | 85 (55.6%) | 232 (65.7%) | .089 | .. | 68 (11%) |
| **Clinical characteristics** | | | | | | |
| Bipolar disorder, type I | 29 (53.7%) | 59 (33.5%) | 148 (38.1%) | .028* | VSA > NVSA, No-SA | - |
| Illness_onset, years‡ | 28.98 (10.35) | 28.35 (12.05) | 32.21 (12.99) | .003** | No-SA > NVSA | 26 (4.2%) |
| Illness duration, years‡ | 28.31 (10.68) | 29.67 (11.59) | 25.65 (13.38) | .005** | NVSA > No-SA | 26 (4.2%) |
| Family history of mental illness | 30 (55.6%) | 115 (65%) | 225 (58.3%) | .255 | .. | 2 (.3%) |
| Family history of suicide | 9 (16.7%) | 35 (19.8%) | 61 (15.7%) | .491 | .. | - |
| Rapid cycling | 10 (24.4%) | 31 (20.1%) | 44 (13.1%) | .047* | VSA, NVSA > No-SA | 89 (14.4%) |
| MADRS score, total‡ | 7.42 (5.82) | 7.32 (5.72) | 6.30 (5.14) | .144 | .. | 11 (1.8%) |
| YMRS score, total‡ | 1.34 (2.64) | 2.33 (3.00) | 2.13 (3.00) | .029 | .. | 13 (2.1%) |
| AIM sum score† | 3.63 (.54) | 3.68 (.60) | 3.51 (.62) | .003** | NVSA > No-SA | 41 (6.6%) |
| PSQI sum score‡ | 6.67 (3.58) | 7.93 (3.53) | 6.79 (3.54) | .001** | NVSA > No-SA | 37 (6.0%) |
| CTQ sum score‡ | 42.06 (11.72) | 44.56 (15.39) | 40.50 (12.47) | .026* | NVSA > No-SA | 29 (4.7%) |
| **Treatment, lifetime** | | | | | | |
| Antipsychotics | 34 (44.4%) | 66 (37.3%) | 149 (38.4%) | .633 | .. | - |
| Lithium | 25 (46.3%) | 58 (32.8%) | 144 (37.1%) | .188 | .. | - |
| Antidepressants | 29 (53.7%) | 86 (48.6%) | 168 (43.3%) | .236 | .. | - |
| Mood stabilizers | 35 (64.8%) | 94 (53.1%) | 218 (56.2%) | .315 | .. | - |
| Sedatives | 13 (24.1%) | 64 (36.2%) | 123 (31.7%) | .230 | .. | - |
| **Comorbidity, lifetime** | | | | | | |
| Anxiety disorder | 13 (25.5%) | 63 (41.2%) | 113 (31.6%) | .039* | NVSA > VSA | 57 (9.2%) |
| Tobacco | 28 (52.8%) | 87 (52.4%) | 177 (48.0%) | .566 | .. | 31 (5.0%) |
| Cannabis | | | | | | 50 (8.1%) |
| Substance use disorder | 16 (31.4%) | 51 (32.5%) | 59 (16.4%) | <.001** | NVSA, VSA > No-SA | 51 (8.2%) |
| Neurological | 17 (34.0%) | 55 (32.4%) | 115 (30.8%) | .870 | .. | 26 (4.2%) |
| Diabetes | 3 (6.3%) | 8 (4.7%) | 30 (8.2%) | .. | .. | 45 (7.2%) |

MADRS, Montgomery-Åsberg Depression Rating Scale(Montgomery and Asberg, 1979); AIM, Affect Intensity Measure (Rubin et al., 2012) ; PSQI, Pittsburg Sleep Quality Index (Buysse et al., 1989); CTQ, Childhood Trauma Questionnaire (Bernstein et al., 2003); YMRS, Young Mania Rating Scale (Young et al., 1978); VSA, Violent suicide attempter; NVSA, Non-violent suicide attempter; No-SA, No suicide attempts.
*p< .05, two-sided, uncorrected; **False Discovery Rate corrected p< .006, two sided.
Post-hoc: Tukey HSD or Games-Howell test according to data normality and homogeneity for continuous variables and $\chi^2$ test for categorical variables. †Normal distribution; ‡Non-normal distribution

Table 2. Metabolic and cognitive factors stratified by suicide attempt history and violence

|  | Violent suicide attempt (N=54) | Non-violent suicide attempt (N=177) | No suicide attempts (N=388) | P-value | $\eta^2$ | Post-hoc | Missing No. (%) |
|---|---|---|---|---|---|---|---|
| **Metabolic health factors** | | | | | | | |
| Body mass index, kg/m$^2$‡ | 27.37 (4.52) | 25.84 (5.13) | 26.98 (4.91) | .016* | .011 | No-SA > NVSA | 23 (3.7%) |
| Waist circumference, cm† | 101.43 (12.75) | 92.89 (12.82) | 97.17 (14.26) | <.001** | .031 | VSA, No-SA > NVSA | 55 (8.9%) |
| Abdominal obesity | 36 (76.6%) | 85 (53.8%) | 197 (54.9%) | .014* | .015 | VSA > NVSA, No-SA | 55 (8.9%) |
| Waist excess | 37 (78.7%) | 99 (62.7%) | 226 (63.0%) | .094 | .008 | VSA > NVSA, No-SA | 55 (8.9%) |
| Insulin resistance – lipid-based measure | 15 (33.3%) | 52 (35.4%) | 130 (39.3%) | .590 | .002 | .. | 87 (14.1%) |
| Insulin resistance – glycemia based measure | 13 (30.2%) | 39 (26.2%) | 97 (29.5%) | .737 | .001 | .. | 98 (15.8%) |
| Metabolic syndrome | 16 (34.8%) | 44 (30.1%) | 133 (39.7%) | .130 | .008 | .. | 92 (14.2%) |
| **Cognitive functions** | | | | | | | |
| Processing speed | | | | | | | |
| - Stroop word condition‡ | 46.26 (7.54) | 48.21 (8.04) | 49.28 (8.44) | .016* | .012 | VSA < No-SA | 120 (19.4%) |
| - Stroop colour condition‡ | 42.04 (9.35) | 44.88 (8.44) | 44.53 (8.28) | .045* | .008 | | 120 (19.4%) |
| - TMT part A‡ | 49.07 (10.72) | 51.72 (10.04) | 50.91 (10.70) | .201 | .004 | .. | 111 (17.9%) |
| - Digit symbol coding‡ | 55.58 (9.96) | 52.78 (9.33) | 53.40 (9.22) | .243 | .006 | .. | 109 (17.6%) |
| - Digit symbol search‡ | 50.72 (9.34) | 49.38 (9.50) | 50.35 (9.77) | .364 | .002 | .. | 108 (17.4%) |
| Executive function | | | | | | | |
| - Colour/word condition of the Stroop test‡ | 47.63 (10.13) | 49.55 (10.38) | 50.48 (9.80) | .342 | .007 | .. | 120 (19.4%) |
| - TMT part B‡ | 46.61 (16.08) | 50.80 (14.74) | 49.21 (14.22) | .348 | .006 | .. | 117 (18.9%) |
| - Semantic verbal fluency† | 42.79 (9.09) | 46.79 (10.83) | 46.42 (10.33) | .073 | .011 | .. | 114 (18.4%) |
| - Phonemic verbal fluency† | 43.89 (10.36) | 48.63 (13.12) | 49.97 (11.97) | .042* | .014 | VSA < No-SA | 117 (18.9%) |
| Verbal memory | | | | | | | |
| - CVLT immediate recall† | 45.02 (11.20) | 51.13 (13.17) | 49.36 (13.47) | .021* | .015 | VSA < NVSA | 107 (17.3%) |
| - CVLT short delay free recall‡ | 42.49 (10.94) | 49.09 (11.38) | 47.29 (13.27) | .006** | .019 | VSA < No-SA, NVSA | 107 (17.3%) |
| - CVLT long delay free recall‡ | 42.08 (11.42) | 47.70 (12.86) | 46.43 (13.48) | .014* | .013 | VSA < NVSA | 108 (17.4%) |

CVLT, California Verbal Learning Test; TMT, Trail Making Test; SA; VSA, Violent suicide attempter; NVSA, Non-violent suicide attempter; No-SA, No suicide attempts.

Abdominal obesity was defined as waist circumference ≥88 cm for females and ≥ 102 for males. Waist excess was defined as waist circumference as excessive when ≥80 cm in females and ≥ 90 in males in normal weight, ≥90 cm and ≥100 cm in overweight, ≥ 105 cm and ≥110 cm in moderately obese, and ≥115 cm and ≥125 cm in severely obese patients.

Insulin resistance - lipid-based: a either triglycerides/high Density Cholesterol ≥1.8 mmol/l or triglycerides ≥1.7 mmol/l or diabetes diagnosis or on lipid lowering medication.

Insulin resistance – glycemia-based: Fasting glycemia ≥ 5.6 mmol/l or diabetes diagnosis.

Post-hoc: Tukey HSD or Games-Howell test according to data normality and homogeneity for continuous variables and $\chi^2$ test for categorical variables.

†Normal distribution
‡Non-normal distribution
*p-value <.05, two-sided.
**FDR-corrected p-value < .006, two-sided.

Table 3. Multiple logistic regression models for the association between visceral adiposity measures and violent SA history

| | Model 1 | | Model 2 | | Model 3: females | |
|---|---|---|---|---|---|---|
| | Exp(B), 95% CI | p | Exp(B), 95% CI | p | Exp(B), 95% CI | p |
| **Probability of violent SA in suicide attempters** | | | | | | |
| **Waist circumference, cm** | 1.039, [1.009-1.069] | .008* | 1.033, [1.002 – 1.065] | .030* | 1.039, [1.004 – 1.076] | .027* |
| **Waist circumference, terciles†** | 1.990, [1.228-3.226] | .005* | 1.984, [1.138 – 3.457] | .010* | 2.068, [1.142 – 3.746] | .004* |
| **Abdominal obesity (sex-based)** | 3.256, [1.455 -7.287] | .005* | 3.029, [1.288 – 7.127] | .005* | 5.184, [1.466 – 18.329] | .012* |
| **Waist excess (sex-based)** | 2.205, [.986 – 4.931] | .052 | 1.773, [.737 – 4.267] | .135 | 2.095, [.731 – 6.008] | .164 |
| **Probability of violent SA in all patients** | | | | | | |
| **Waist circumference** | 1.026, [1.003 – 1.049] | .019* | 1.023, [.998 – 1.048] | .018* | .058, [-.004 - .053] | .058 |
| **Waist circumference, terciles*** | 1.666, [1.094 – 2.537] | .005* | 1.659, [1.069 – 2.574] | .008* | 1.896, [1.083 – 3.320] | .017* |
| **Abdominal obesity (sex-based)** | 2.579, [1.273 – 5.225] | .006* | 2.157, [1.003 – 4.641] | .009* | 5.374, [1.574 – 18.351] | .005* |
| **Waist excess (sex-based)** | 2.107, [1.015 – 4.373] | .040* | 1.672, [.781 – 3.577] | .059 | 1.916, [.699 – 5.247] | .164 |

Variables not demonstrated in the table in Models 1: sex and bipolar disorder type; Model 2: sex, bipolar disorder type, and anxiety disorder history; Model 3: bipolar disorder type.

†In models 1 and 2 whole sample waist circumference terciles were used. In model 3, female terciles were used.

Significance level was set at p<.05, two-sided. P-values were obtained using Bootstrapping with 1000 imputations.

Figure 1. Probability of suicide attempt and violent suicide attempt in all patients, suicide attempters, and female suicide attempters

Violent SA probability in all patients

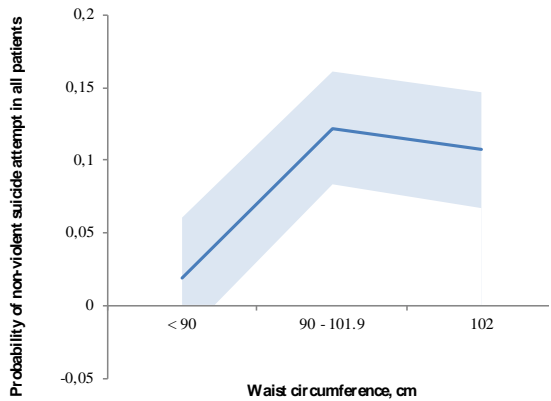

Violent SA probability in suicide attempters

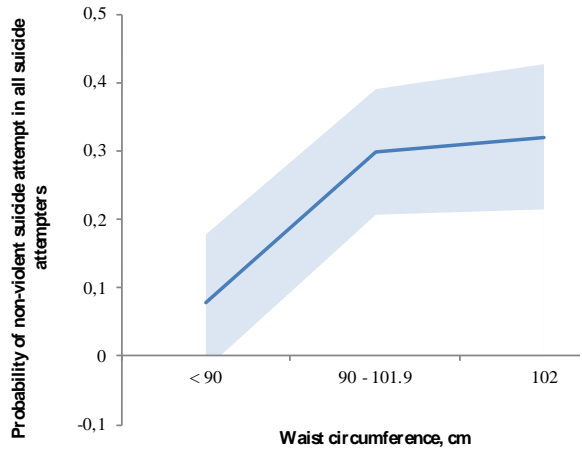

Violent SA probability in all females

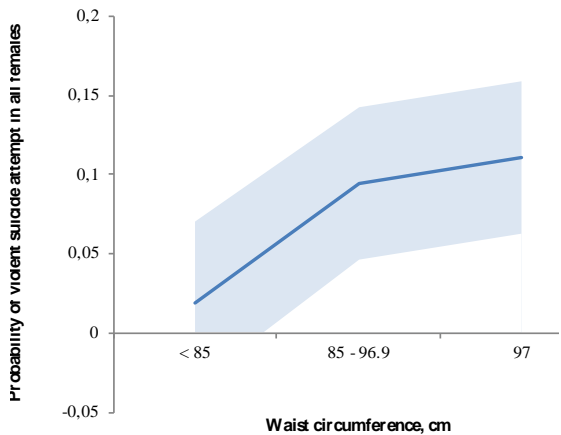

Violent SA probability in female suicide attempters

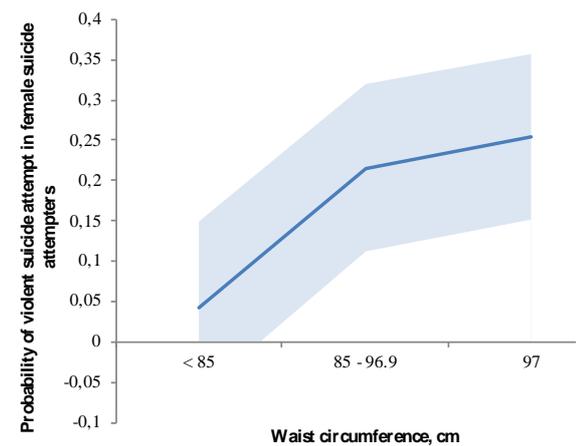

The first plot shows estimated marginal means with 95% confidence intervals of violent and non-violent SA proportionally to all patients (N = 564). Statistical test was a linear model with sex as covariate (sex = .40), p = .001*. The second plot shows estimated marginal means with 95% confidence intervals of violent SA proportionally to all SA (N = 205). Statistical test was a linear model with sex as covariate (sex = .27), p = .001**. The third plot shows estimated marginal means with 95% confidence intervals of violent SA proportionally to all females (N = 339), p = .025*. The fourth plot shows estimated marginal means with 95% confidence intervals of violent SA proportionally to all SA in females (N = 142). Statistical test was a linear model, p = .013*. Only cases with complete data were analyzed. p-values were obtained by Bootstrapping (1000 imputations). Differences in males were not significant and are not represented here.

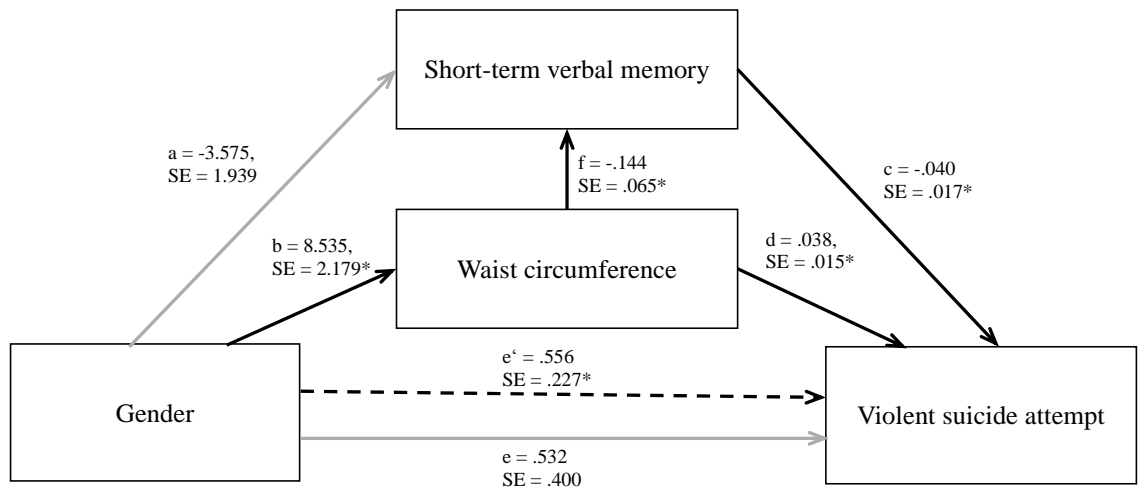

Figure 2. The applied mediation model between sex and violent suicide attempt via waist circumference and short-term verbal memory. Unstandardized coefficients and standard errors (SE) for each path of the mediation model. To obtain f and correct a and b values, two separate mediation models were carried with violent suicide attempt as an outcome variable, sex as a covariate and either waist circumference as an independent variable and short-term verbal memory as a mediating variable or vice-versa. e represents the direct effect and e' the indirect effect. *p < 0.05, with Bootstrapping with 1000 iterations.